\documentclass{ws-procs975x65}

\usepackage{color}

\begin{document}

\title{METHOD OF MATCHED EXPANSIONS \& \\ THE SINGULARITY STRUCTURE OF THE GREEN FUNCTION
}

\author{MARC CASALS$^1$, SAM DOLAN$^2$, ADRIAN C. OTTEWILL$^3$ and BARRY WARDELL$^4$}

\address{$^1$School of Mathematical Sciences, Dublin City University, Glasnevin, Dublin 9, Ireland. E-mail: marc.casals@dcu.ie
\\
$^2$School of Mathematics, University of Southampton, University Road, Southampton, SO17 1BJ, 
United Kingdom.
E-mail: s.dolan@soton.ac.uk 
\\
$^3$Complex and Adaptive Systems Laboratory and School of Mathematical Sciences,
University College Dublin, Belfield, Dublin 4, Ireland.  E-mail: adrian.ottewill@ucd.ie
\\
$^4$Max-Planck-Institut f\"ur Gravitationsphysik, Albert-Einstein-Institut, 14476 Golm, Germany.  E-mail: barry.wardell@aei.mpg.de}

\begin{abstract}
We present the first successful application of the method of Matched Expansions for the calculation of the self-force on a point particle in a curved spacetime. We investigate the case of a scalar charge in the Nariai spacetime, which serves as a toy model for a point mass moving in the Schwarzschild black hole background. We discuss the singularity structure of the Green function beyond the normal neighbourhood and the interesting effect of caustics on null wave propagation.
\end{abstract}


\bodymatter

\section{Introduction}\label{Intro}

Extreme Mass Ratio Inspirals, such as a stellar-mass compact object orbiting around a 
supermassive black hole,
can be understood within perturbation theory of General Relativity 
as the smaller body's motion deviating from the background geodesic due to a {\it self-force}~\cite{Poisson:2003}. 
The calculation of the self-force is necessary
in order to model the orbital motion and obtain accurate gravitational-wave templates.
The expression
for the self-force in a general
spacetime is formally similar for the cases of an orbiting point scalar charge,
electrical charge and a point mass.
In all these cases, the self-force is given by some local
terms which are relatively 
easy
to calculate, plus
a {\it `tail'} term which, in the case of a scalar charge $q$ following a worldline $z(\tau)$, is given by
\begin{equation} \label{eq:S-F}
q^2\int_{-\infty}^{\tau^-} \nabla_\mu G_{ret} (z(\tau), z(\tau')) d \tau'
\end{equation}
where $G_{ret}(x, x')$ 
is the {\it `retarded' Green function} 
of the scalar field wave equation.
As 
the `tail' term depends on the {\em entire} past history of the particle's worldline,
it is usually  quite difficult to calculate and one has to do so numerically. 

\section{Method of Matched Expansions}
We
choose to 
use the method of Matched Expansions~\cite{Poisson:Wiseman:1998} to calculate (\ref{eq:S-F}).
In this method, the `tail' integral is separated into two regimes:

(a) In the {\it quasilocal} regime, 
$\tau' \in (\tau-\Delta \tau, \tau)$
for a certain `small value' $\Delta\tau$, the Green function is calculated using
the {\it Hadamard form}:
\begin{equation} \label{eq:Hadamard}
G_{ret}\left( x,x' \right) = \theta_{-} \left( x,x' \right) \left\lbrace \Delta^{1/2} \left( x,x' \right) \delta \left( \sigma\right) - 
V \left( x,x' \right) \theta \left( - \sigma  \right) \right\rbrace
\end{equation}
where $\sigma$ is 
half the square of the geodesic distance
 between the spacetime points $x$ and $x'$, 
the {\it van Vleck determinant} $\Delta(x,x')$ 
 and $V(x,x')$ are two regular biscalars
and $\theta_{-} \left( x,x' \right)$ guarantees
causality.
Given the structure of (\ref{eq:Hadamard}) it is clear that only $V(x,x')$ will contribute to the `tail' integral in the quasilocal regime.
The Hadamard form is only valid for points `nearby' - within a {\it normal neighbourhood}.

(b) In the {\it distant past} regime,  
$\tau' \in (-\infty, \tau-\Delta\tau )$,
the Green function can be calculated using a Fourier mode decomposition
and
deforming
the frequency-integral into a contour in the complex frequency-plane.

\section{Singularity structure of the Green function}
We carried out~\cite{CDOWa} the first succesful application of the method of Matched Expansions.
With the future view of applying it to the case of a point mass moving in an astrophysically-relevant spacetime, we used
a toy model where calculations are easier:
 a scalar charge moving in the Nariai spacetime
 (cartesian product of 2-dimensional de Sitter spacetime and the 2-sphere),
 in which the radial potential of Schwarzschild is replaced by the P\"oschl-Teller potential.

In the quasilocal regime, some powerful methods have been developed~\cite{CDOWb,Ottewill:2009uj}
for the calculation of $V(x,x')$ up to, and beyond,
the normal neighbourhood.

\begin{figure}[h!]
\begin{center}
\psfig{file=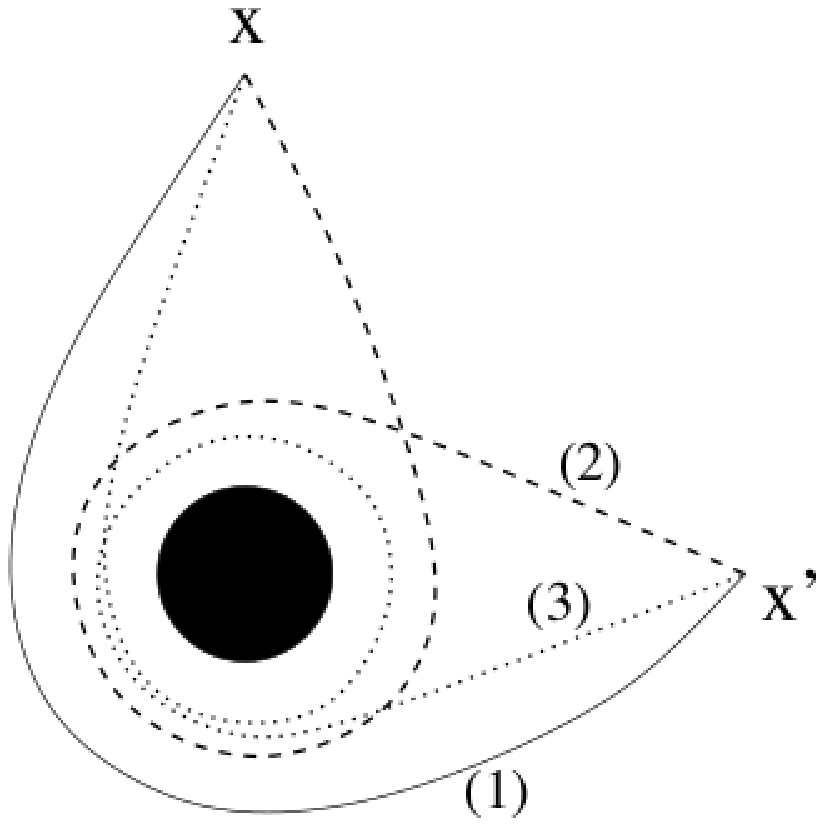,width=4cm}
\psfig{file=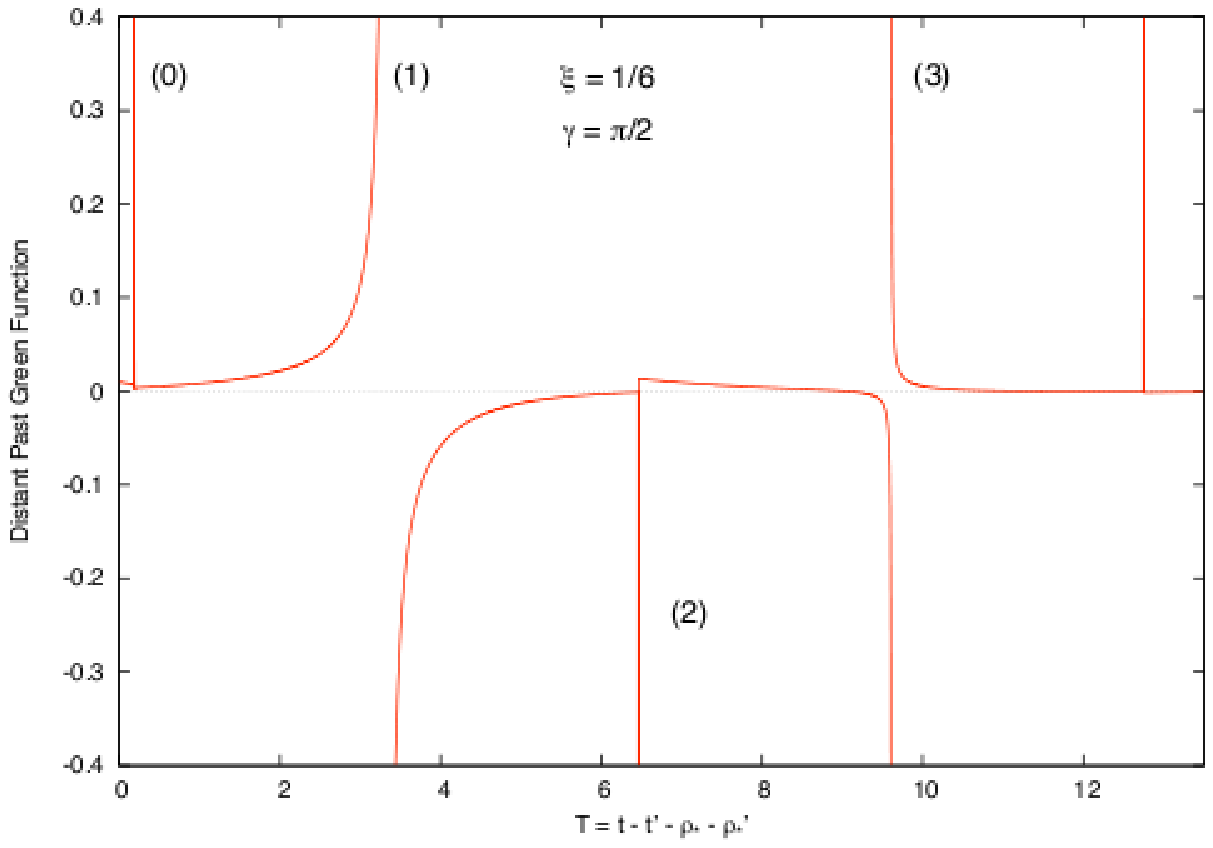,width=6cm}
\end{center}
\caption{
(a)
Null geodesics orbiting the unstable photon orbit of the background.
(b)
`Retarded' Green function as a function of the proper time of the 
static
 scalar charge in Nariai.
There is a
1-to-1
correspondence between the
geodesics and the
times when the Green function becomes singular,
with a four-fold structure: $\delta(\sigma),\ \frac{1}{\pi\sigma},-\delta(\sigma),\ -\frac{1}{\pi\sigma},\ \delta(\sigma),\dots$
}
\label{fig:geodesics}
\end{figure}

In the Nariai spacetime, the `retarded' Green function 
in the `distant past' is completely given by the sum over its poles - the {\it quasinormal modes}.
 A study of this series beautifully uncovered -- Fig.\ref{fig:geodesics} -- the 
singularity structure of the Green function: its singularities occur
at the times 
it takes
a null geodesic
to travel
between $x$ and $x'$ (i.e., when $\sigma=0$); 
furthermore, the singularity structure follows a 4-fold cycle: $\delta(\sigma), 1/(\pi\sigma), -\delta(\sigma), -1/(\pi\sigma), \delta(\sigma),\dots$.

An alternative, heuristic way for understanding how this structure arises is by use of the {\it Feynman Green function}, 
which has Hadamard form
\begin{equation} \label{eq:Hadamard G_F}
G_F\left( x,x' \right) =\frac{i}{2\pi}  \left\lbrace \frac{\Delta^{1/2}  \left( x,x' \right)}{ \sigma+i\epsilon}+ 
V \left( x,x' \right) \ln\left( \sigma+i\epsilon \right)+W(x,x') \right\rbrace
\end{equation}
where $W(x,x')$ is another regular bitensor and $\epsilon>0$. The `retarded' Green function can be obtained as
$G_{ret}(x,x')=2\theta_-(x,x')\lim_{\epsilon\to 0}\text{Re}\left(G_F(x,x')\right)$.
Although strictly speaking the Hadamard form is only valid within the normal neighbourhood,
if we tentatively use it beyond 
 we can obtain the van Vleck determinant from a transport equation past caustic points, where it diverges.
One then finds that 
$\Delta^{1/2}  \left( x,x' \right)$
picks up a factor `$-i$'  everytime a null geodesic crosses a caustic.
From the property $\lim_{\epsilon\to 0^+}i/( \sigma+i\epsilon)=\pi\delta(\sigma)+i/\sigma$, one can see how the 4-fold cycle arises.
This factor seems to be typical of the $\mathbb{S}^2$ topology and,
although it is known in other fields of Science, in General Relativity
it has only been previously observed by Prof.~Ori~\cite{Ori1short}, to the best of our knowledge.

\begin{figure}[h!]
\begin{center}
    \psfig{file=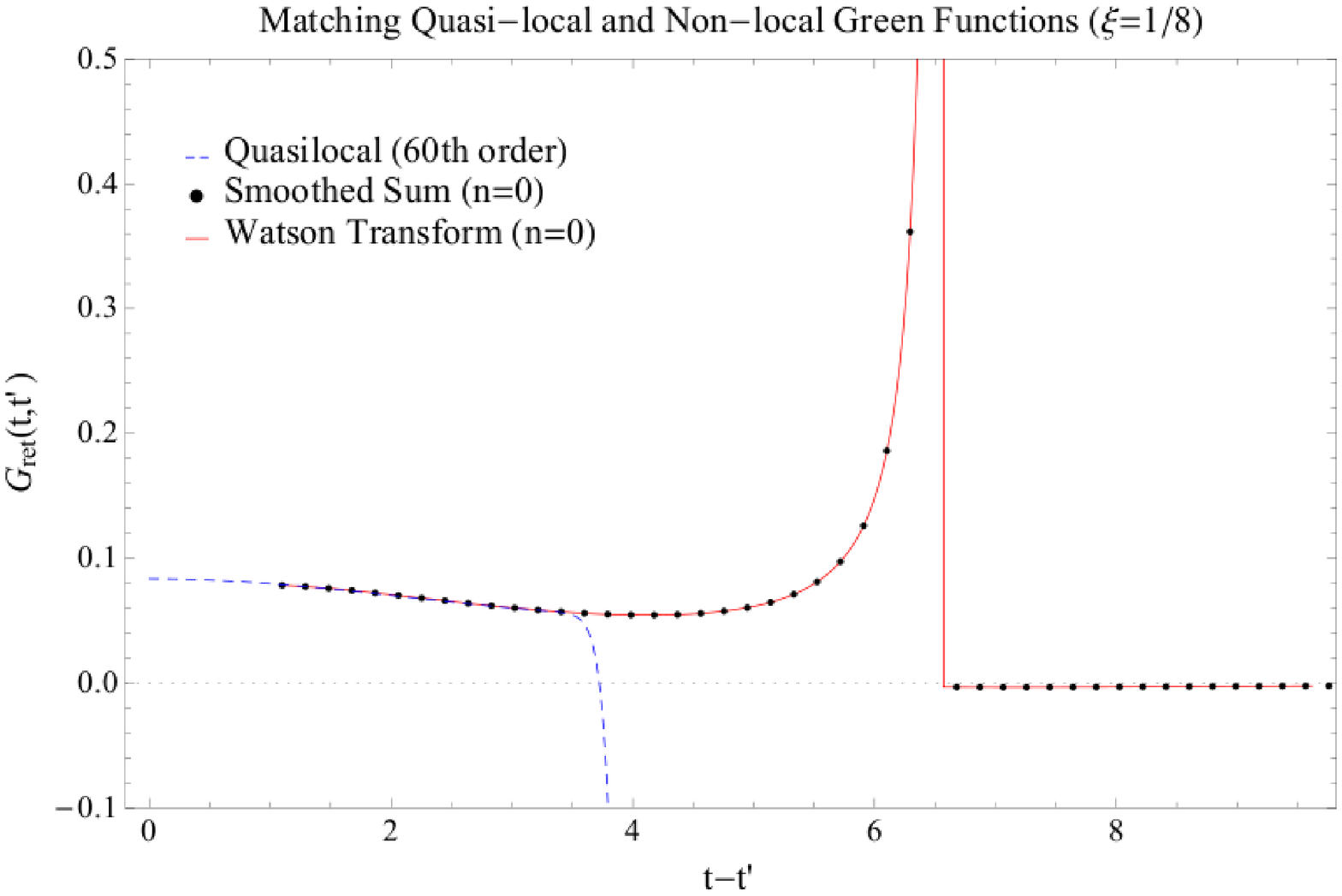,width=6cm}
        \psfig{file=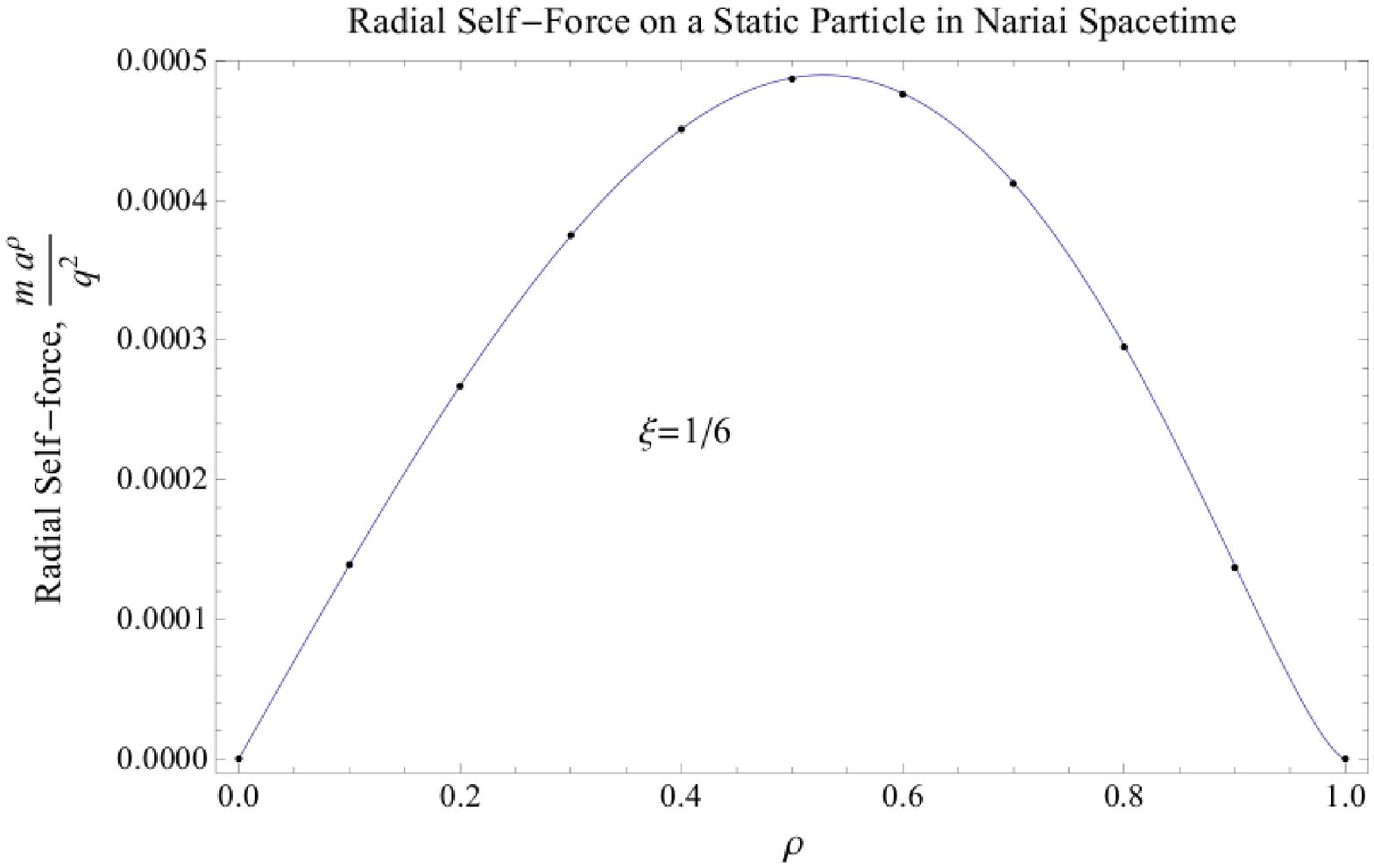,width=6.5cm}
\end{center}
\caption{
(a) Matching of the `quasilocal' and `distant past' contributions
as a function of time, 
(b) self-force for a static scalar charge in the Nariai spacetime as a function of the radius.
}
\label{fig:matching}
\end{figure}

We were able to successfully match the quasilocal and `distant past' contributions to the `retarded' 
Green function -- Fig.\ref{fig:matching}(a).
The calculation of the `retarded' Green function over the whole past of the particle's worldline enabled us to
calculate the self-force for a scalar charge in Nariai -- Fig.\ref{fig:matching}(b).

\section{Acknowledgements}
M.C. and B.W. are supported by the Irish Research Council for Science, Engineering and Technology.


\bibliographystyle{ws-procs975x65}
\bibliography{main}

\end{document}